\documentclass[final,5p,times,compress,twocolumn]{elsarticle}

\usepackage{amssymb}
\usepackage{amsmath}
\usepackage{lipsum}
\usepackage{xcolor}
\usepackage{booktabs}
\usepackage{hyperref}
\usepackage{orcidlink}
\usepackage[normalem]{ulem}
\usepackage{enumitem}
\usepackage{lineno}

\journal{Physics Letters B}

\begin{document}

\begin{frontmatter}

\title{Signatures of $K^-$ condensation on neutron star structure and $f-$mode frequencies}

\author[BITS_Hyd]{Debanjan Guha Roy\orcidlink{0009-0004-5222-6690}*}
\author[BITS_Hyd]{Sarmistha Banik\orcidlink{0000-0003-0221-3651}}
\cortext[cor1]{Corresponding author: debanjan.gr@gmail.com}
\affiliation[BITS_Hyd]{organization={Department of Physics, BITS Pilani, Hyderabad Campus},%Department and Organization
            % addressline={}, 
            city={Hyderabad},
            postcode={500078}, 
            state={Telengana},
            country={India}}

\begin{abstract}
%% Text of abstract
Antikaon ($K^-$) condensation within neutron star matter (NS) depends on the antikaon-nucleon interaction potential ($U_K$). Appearance of $K^-$ generally softens the equation of state (EOS). The impact of this softening on the structure of the NS can be leveraged to find a telltale sign of the phase transition from nucleonic matter to $K^-$ condensation.
To investigate the impact of $K^-$ condensation on NS properties using a Bayesian inference framework, we choose two sets of RMF model parameters to obtain a stiff (DD2) and relatively soft (FSU) nucleonic EOS, and explore a wide range of optical potential depths.
Multimessenger observations from NICER and LIGO/Virgo constrain the optical potential values to $U_K = -104.72^{+13.82}_{-12.48}$ MeV and $U_K = -66.46^{+2.47}_{-3.42}$ MeV for the stiff and soft cases, respectively.
Deeper $K^-$ potentials trigger condensation at a lower density, softening the EOS and lowering the corresponding maximum masses.
While slopes of mass-radius and tidal deformability curves overlap between nucleonic and exotic EOSs, their curvature and $f-$mode oscillation properties (frequency and damping time) reveal features attributable to EOS softening. However, distinguishing the specific exotic degrees of freedom responsible for the softening remains an open challenge.
\end{abstract}


\end{frontmatter}

%\tableofcontents

% \linenumbers

%% main text

\section{Introduction}
\label{sec:intro}

At the extreme densities encountered in the cores of neutron stars, the emergence of exotic forms of matter beyond the conventional nuclear composition becomes possible\cite{glendenning}. One such candidate is the antikaon ($K^-$) condensate. The threshold density for the onset of $K^-$ condensation is highly sensitive to the depth of the antikaon optical potential ($U_K$). A deeper potential results in the earlier appearance of $K^-$ mesons, leading to a significant softening of the equation of state (EOS), which in turn reduces the maximum mass supported by the star \citep{Malik_ApJ_2021, Banik_PRC2001}. 

However, the value of the in-medium $K^-$-nuclear potential remains a key open question. While $U_K$ derived from chirally motivated amplitudes at threshold is relatively shallow, phenomenological studies of kaonic atoms indicate much deeper, strongly attractive potentials. Calculations suggest depths of about 80-90 MeV for kaonic atoms \cite{Friedman1999} and up to 100-110 MeV for antikaon-nuclear quasibound states\cite{Friedman2007}. Recent kaonic-hydrogen X-ray data from SIDDHARTA \cite{Bazzi2012}, together with earlier KEK E228 \cite{KEKE228_Iwasaki1995} and DEAR \cite{DEAR_Beer2005} measurements, provide crucial constraints on the subthreshold $\bar{K}N$ amplitude. The ongoing SIDDHARTA-2 kaonic-deuterium experiment \cite{Sirghi2024}, combined with these results, is expected to further clarify the strength of the $\bar{K}N$ interaction.

Recent observational breakthroughs---particularly from gravitational wave (GW) detections and high-precision pulsar measurements---have significantly advanced our understanding of the internal structure of NSs \cite{Raaijmakers2021, Rutherford2024, Margalit2017, Radice2018, Tews_PRC2018}. The Neutron Star Interior Composition Explorer (NICER) has provided simultaneous mass and radius measurements with unprecedented accuracy. For instance, NICER observations of PSR J0030$+$0451 yielded an estimated radius of $12.71^{+1.14}_{-1.19}$ km and a mass of $1.34^{+0.15}_{-0.16}~M_\odot$ \cite{Riley_2019}, while an independent analysis reported a radius of $13.02^{+1.24}_{-1.06}$ km and a mass of $1.44^{+0.15}_{-0.14}~M_\odot$ for the same source~\cite{Miller_2019}. Additionally, NICER's measurements for the more massive pulsar PSR J0740$+$6620 indicate a radius of $12.39^{+1.30}_{-0.98}$ km and a mass of $2.072^{+0.067}_{-0.066}~M_\odot$, within a 68\% confidence interval \cite{Riley_2021}. Most recently, NICER observations of PSR J0437$-$4715 determined its mass and radius to be $1.418^{+0.037}_{-0.037}~M_\odot$ and $11.36^{+0.95}_{-0.63}$ km, respectively \cite{Choudhury_2024}. 
Current multimessenger observations place significant bounds on the dimensionless tidal deformability ($\Lambda$) of NSs. From GW170817 alone,  LIGO/Virgo report $\Lambda_{1.36} = 190^{+390}_{-120}$ at 90\% credibility~\cite{LIGOScientific:2018cki}.
Assuming purely hadronic EOSs, some studies obtain $\Lambda_{1.4} \geq 120$ \cite{Annala_PRL2018}.

Phase transitions from nuclear matter to exotic forms—such as those involving antikaons, hyperons, or deconfined quarks are expected to leave imprints on the $M-R-\Lambda$ relation \cite{Chatterjee_EPJA2016, Tolos_PPNP2020, Annala:2019puf}. One way to identify them is through studying the slope of the $M-R$ relation or the $M-\Lambda$ relation. Ref. \cite{SunLattimer2025, Ferreira_PRD2024} shows how including the slope $dR/dM$ improves the accuracy of their method of determining the EOS by inverting the $M-R$ relation. In a very recent study \cite{Bauswein_curvature_2025}, curvature of the curves in the $R(M)$ domain has been used to distinguish EOSs with hyperons from purely nucleonic ones.

Complementary to such bulk-parameter analyses, oscillation modes of NSs—particularly the fundamental or $f-$modes excited during the ringdown phase following a binary NS merger are strongly dependent on the EOS of the interior\cite{ZhaoLattimer2022}. The $f-$modes correspond to global fluid oscillations restored by pressure, typically with a frequency $\sim 2$ kHz \cite{Lindblom_1983, Andersson_1998}, and are strongly coupled to gravitational wave (GW) emission. Their frequencies and damping times follow quasi-universal relations with mass, radius, and compactness, thereby offering sensitive probes of possible exotic phases in dense matter \cite{Pradhan_PRC2022, Sotani_PRD2011, Thakur_PRD2024, Kumar_EPJC2024, Kumar_JCAP2023}.
Future GW detectors like Cosmic Explorer (CE) \cite{CE_WhitePaper2022} and Einstein Telescope (ET) \cite{ET_WhitePaper2020}, with enhanced sensitivity to post-merger signals and NS oscillations, could provide additional, independent constraints on the EOS.

In this work, we aim to investigate how deep the antikaon potential can be while still allowing the maximum NS mass to remain consistent with observational constraints. For the first time, we calculate fundamental mode ($f-$mode) frequencies and their damping times for NSs with $K^{-}$ condensates inside their core.
Furthermore, we analyse the curvature of the $R(M)$ curve and first and second-order derivatives of the radius, tidal deformability and $f-$mode frequency expressed as functions of mass, to probe potential signatures of $K^{-}$ condensates in NSs.
We find that deeper $K^{-}$ potentials trigger earlier condensation, softening the EOS and lowering the maximum mass with associated radii depending on the stiffness of the nucleonic EOS. Moreover, curvature and second derivatives can help us distinguish between EOSs with nucleonic and kaonic degrees of freedom.

We lay out the formalism for the EOSs with antikaon condensation, calculation of stellar observables, and our Bayesian inference setup in the next section. The results are described in Sec. \ref{sec:Results} followed by the conclusions in Sec. \ref{sec:Conclusions}.

\section{Formalism}
\label{sec:Formalism}
\subsection{Equations of State}
In this paper, the Relativistic Mean Field (RMF) model \cite{glendenning} is employed to describe dense matter in the core of NS, which consists of neutrons, protons, electrons and muons. We consider two different sets of model parameters: DD2 \cite{Hempel_NucPhysA2010} and FSU \cite{Bunta_PRC2003}, both consistent with NICER and GW observational data \cite{Nandi_2019}, DD2 being slightly stiffer than FSU. 
In the DD2 model, the meson couplings are density-dependent, while in FSU-type models, an additional nonlinear $\omega$ coupling is considered to take care of the high-density behaviour of the NS core. The $K^-$ condensates are treated in the RMF formalism as in \cite{Banik_thirdFAMILY2001, BanikChar_PRC2014, Malik_ApJ_2021}. Experimentally, the precise value of the attractive $U_K$ in nuclear matter remains uncertain.  To account for this, we construct the EOSs over a wide range of $U_K$, spanning $-60~\text{MeV}$ to $-180~\text{MeV}$.
This range is consistent with the unitary chiral model calculations, phenomenological fit to kaonic atom data \cite{Friedman1994, Tolos_PRC2006, Tolos_PPNP2020}, and some recent works---a Bayesian analysis of $K^-$ condensation \cite{Parmar2024} and a study of radial oscillation of NSs with antikaon condensates \cite{Kheto2023}.
\subsection{Neutron Star Observables}
The EOS sets are used to calculate the static structure parameters (masses $M$, radii $R$)  of the spherically symmetric stars described by the Tolman-Oppenheimer-Volkoff  (TOV) equations.  
The dimensionless tidal deformability ($\Lambda$) is sensitive to the star radius and also depends on the EOS of NS matter. These calculations involve solving an additional first-order differential equation in parallel with the TOV equations \cite{Postnikov2010}.

Next, we compute the $f-$mode frequency and the associated damping time of quasinormal oscillations in NS with antikaon condensates, within the framework of linearised general relativity \cite{GuhaRoy_ApJ2024, GuhaRoy_PLB2024}. These stellar oscillations carry imprints of the internal structure, with the real part of the QNM frequency corresponding to the oscillation frequency and the imaginary part representing the exponential damping rate.

Finally, we do a similar analysis as in Ref. \cite{Bauswein_curvature_2025} to explore the effects and characteristic signatures of  $K^-$ condensation on the curvature and the second derivative of radius, tidal deformability, $f-$mode oscillation frequency with respect to stellar mass in order to quantitatively distinguish exotic EOSs from purely nucleonic EOSs. 

\subsection{Bayesian Analysis}
\label{sec:Bayes}
We employ a Bayesian framework to infer the EOS parameters---the RMF parameters are fixed as per DD2(FSU) nucleon model, while $U_K$ is sampled uniformly over $[-180, -60]\,\mathrm{MeV}$. The corresponding model with $K^-$ condensation is identified hereafter as the DD2+$K^-$(FSU+$K^-$) setup.

We constrain the EOS models using multiple astrophysical observations:\\
(i) NICER mass–radius data for PSR J0030+0451 \cite{Riley_2019, Miller:2019cac} and PSR J0740+6620 \cite{Riley:2021pdl, Miller:2021qha}, \\
(ii) tidal deformability of a $1.36$ $M_\odot$ star from GW170817 \cite{LIGOScientific:2018cki}, and \\
(iii) lower limits on the maximum mass from Shapiro delay measurements of massive pulsars (PSR J0740$+$6620, PSR J0348$+$0432 \cite{Antoniadis2013}, PSR J1614$-$2230 \cite{Arzoumanian2018}) as in Ref. \cite{Rezzolla_Jiang_2023}.
\begin{figure}
    \centering
    \includegraphics[width=\linewidth]{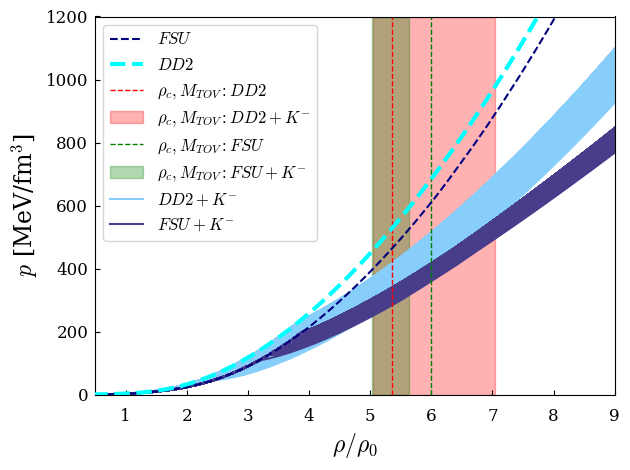}
    \caption{EOS corresponding to $U_K$ values from the posterior samples in the pressure ($p$) - baryon number density ($\rho$) plane. $\rho_0=0.16\text{ fm}^{-3}$ is the number density at nuclear saturation. Appearance of kinks signals phase transition due to $K^-$ condensation.}
    \label{fig:eos_from_posterior}
\end{figure}
The mass–radius probability distributions for calculating the likelihood are obtained using kernel density estimation (KDE) \cite{TerrellScott1992} with a Gaussian kernel applied to the publicly available posterior samples for PSR J0030+0451 and PSR J0740+6620.
Sampling is carried out with \texttt{PyMultiNest} \cite{Buchner_2014} using 2000 live points in each setup, parallelised with OpenMPI to accelerate computations across multiple CPUs. For each setup, we obtain approximately 4000 posterior EOSs featuring $K^-$ condensates.

\section{Results}
\label{sec:Results}

\subsection{Equations of State}

Fig.~\ref{fig:eos_from_posterior} shows the pressure–baryon number density relation for two RMF models: DD2 (thick dashed, light blue) and FSU (thin dashed, blue) with nucleons and leptons only; DD2 is slightly stiffer. Shaded bands indicate cases with antikaons for varying $U_K$, where deeper $U_K$ leads to $K^-$ onset at lower densities and shallower $U_K$ shifts it higher. The kinks mark the threshold where the energy of $K^-$ mesons $\omega_{K}$ exceeds the electron chemical potential $\mu_{e}$, signaling $K^-$ condensation; threshold densities for each model and $U_K$ are listed in Tab.~\ref{tab:thres} (e.g., DD2 (FSU) allows onset at $2.32 (3.56)~\rho_0$ for $U_K \sim -180$ MeV ($\sim -90$ MeV), where $\rho_0=0.16 fm^{-3}$ is the baryon number density at nuclear saturation). The red (green) bands show the range of central densities $\rho_c = 5.05 - 7.03~\rho_0$ ($5.03 - 5.64~\rho_0$) for DD2+$K^-$(FSU+$K^-$); the lower and median values nearly coincide, while the upper limit is larger for the DD2+$K^-$ case(See Tab. 2). Vertical dashed lines mark the nucleon-only central densities at $5.6\rho_0$ (DD2) and $6.0\rho_0$ (FSU). 

\begin{table}
    \centering
    \begin{tabular}{c c c c}
        \toprule
         EOS set & maximum & median & minimum\\
         \midrule
         DD2+$K^-$ & 4.11 & 3.33 & 2.32\\
         FSU+$K^-$ & 4.11 & 3.99 & 3.56\\ 
         \bottomrule         
    \end{tabular}
    \caption{Threshold number density, $\rho_{thres}$ for equations of state with $K^-$ condensates in multiples of saturation number density, $\rho_0=0.16$ fm$^{-3}$}
    \label{tab:thres}
\end{table}
\begin{table}
    \centering
    \begin{tabular}{c c c c}
        \toprule
        EOS set & maximum & median & minimum  \\
        \midrule
        DD2+$K^-$ & 7.03 & 5.18 & 5.05\\
        FSU+$K^-$ & 5.64 & 5.17 & 5.03\\ 
        \bottomrule         
    \end{tabular}
    \caption{Central baryon number density $\rho_{c}$ for maximum mass neutron star $M_{TOV}$ with $K^-$ condensates in multiples of $\rho_0=0.16$ fm$^{-3}$.}
    \label{tab:number_density}
    \end{table}

\begin{figure}[t]
    \centering
    \includegraphics[width=\linewidth, height=6cm]{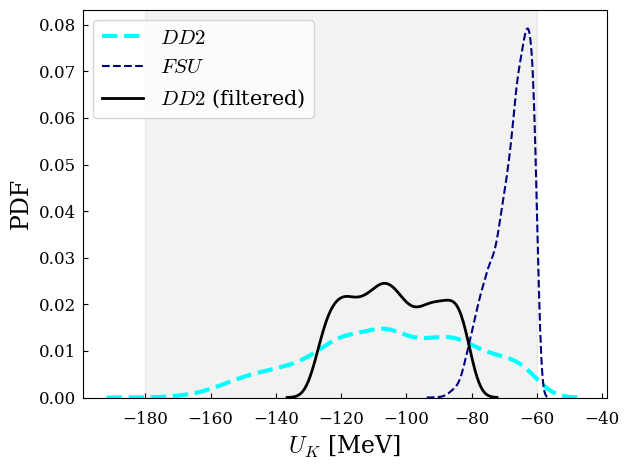}
    \caption{Probability density function (PDF) of $U_K$ from the Bayesian posterior. Thick (thin) dashed light blue (blue) curves show the kernel density estimates for DD2 (FSU), while the shaded rectangle indicates the uniform prior range. The solid black line marks the result of imposing additional constraint on maximum mass and radius at maximum mass \cite{Fan2024}.}
    \label{fig:U_K_PDF}
\end{figure}

\begin{figure*}[t]
    \centering
    \includegraphics[width=0.9\linewidth]{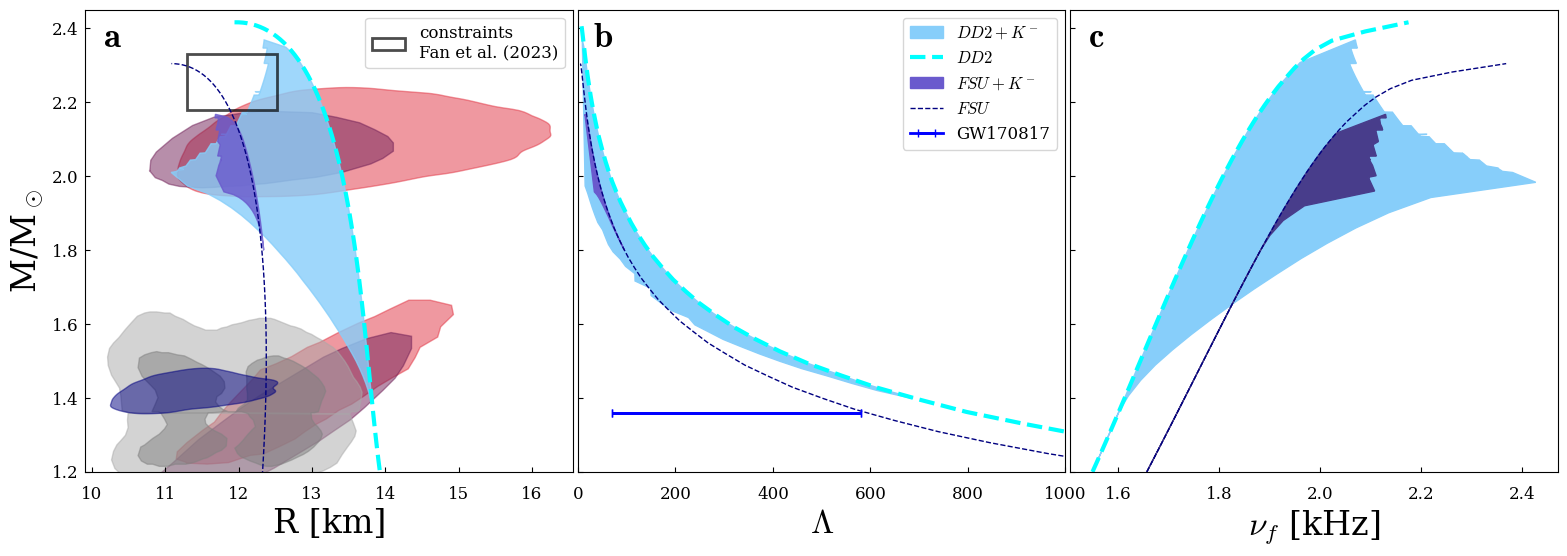}
    \caption{\textbf{Panel a}: Mass-radius relation ($M-R$) for the equations of state from the posterior samples. The grey region shows the joint posterior distribution (dark grey - 50\% credible interval and light grey - 90\% credible interval) of mass-radius values for the two components of the binary neutron star merger event GW170817. The \textbf{blue} patch around 1.4 $M_\odot$ represents 68\% credible interval for one of the most recent NICER observations: PSR J0437$-$4715\cite{Choudhury_2024}. The rest of the patches illustrate the 68\% credible interval \textbf {(dark - Riley, light - Miller)} from the NICER observation data for the two pulsars: PSR J0030$+$0451 (patches around 1.4 $M_\odot$) and PSR J0740$+$6620 (patches around 2.0 $M_\odot$). \textbf{Panel b}: Mass-tidal deformability ($ M-\Lambda$) for the EoS from the posterior samples. The horizontal bar shows the constraints on the tidal deformability of a 1.36 $M_\odot$ neutron star from the GW170817 event. \textbf{Panel c}: Mass-frequency ($M-\nu_f$) of $f-$modes for the EOS from the posterior samples.}
    \label{fig:MRLamNu}
\end{figure*}

Fig.~\ref{fig:U_K_PDF} shows the posterior probability distribution of the optical potential $U_K$ (in MeV) from our Bayesian analysis. Dashed and dotted curves denote the kernel density estimates for the DD2+$K^-$ and FSU+$K^-$ cases, respectively, over the assumed uniform prior range ($-180$ to $-60$ MeV, shaded). For the given set of astrophysical observations, antikaon condensation occurs for the DD2 model with a median value of $U_K = -104.72^{+13.82}_{-12.48}$ MeV with the associated 68\% credible interval (CI). In contrast, the softer FSU model favours $K^-$ condensation only within a narrower band: $U_K = -66.46^{+2.47}_{-3.42}$ MeV.
Imposing the maximum mass constraints as reported in the Ref. \cite{Fan2024} (see rectangular box in the left panel of Fig. \ref{fig:MRLamNu}) further restricts $U_K$ values ($104.58^{+8.65}_{-7.55}$) MeV for the DD2+$K^-$ case, as marked by the solid black line. 

\begin{table}
    \centering
    \begin{tabular}{c c c c}
        \toprule
        EOS set & maximum & median & minimum \\
        \midrule
        DD2+$K^-$ & 2.37 (12.35) & 2.26 (12.34) & 2.00 (11.08)\\
        FSU+$K^-$ & 2.17 (11.68) & 2.14 (11.71) & 2.00 (11.69)\\
        \bottomrule         
    \end{tabular}
    \caption{Maximum mass $M_{TOV}$ in $M_\odot$ units and the corresponding radius (R in km)  for a given EOS.}
    \label{tab:M-R}
\end{table}

\subsection{Neutron Star Observables}
In Fig. \ref{fig:MRLamNu} we plot mass ($M$) as a function of radius ($R$), dimensionless tidal deformability ($\Lambda$), and $f-$mode frequency ($\nu_f$) in the three panels respectively.
The deep blue and light blue dashed lines represent the nucleon-only cases, while the colored patches correspond to the posterior samples including $K^-$ condensation. 

In the left panel, i.e. Fig. 3a, the maximum masses (radii) for nucleons-only cases are $2.42~M_\odot$($11.94$ km) and  $2.30~M_\odot$($11.08$ km) for the models DD2 and FSU, respectively.
The presence of $K^-$ condensates lowers the maximum masses within the range: $2.00$ - $2.37$ $M_{\odot}$, and the radii for maximum masses shrink and lie between $11.08$ and $12.41$ km for the DD2 model for the range of $U_K$. The maximum masses for the FSU model vary from $2.0$ to $2.17$ $M_{\odot}$. Their radii, however, show a minimal variation with a median of $11.72$ km, ranging from $11.68$ to $11.77$ km. The values are listed in Tab. \ref{tab:M-R}.

Examining the maximum mass and the corresponding radius as they vary with $U_K$ for the DD2+$K^-$ case (light blue patch), we notice that the variation is non-linear. This feature is observed for the FSU+$K^-$ case (blue patch) as well. The exact trend for the variation, however, depends on the stiffness of the underlying nucleonic EOS.

The surrounding shaded patches in Fig. 3a indicate observational constraints from various astrophysical measurements:
mass-radius inferences by NICER from X-ray observations of: the high-mass pulsar PSR J0740$+$6620 and the low-mass pulsar PSR J0030$+$0451 (in two different shades of pink), and PSR J0437$-$4715 (blue).
The grey patches mark the mass-radius inferences from the gravitational wave data for the binary NS merger event GW170817.
The DD2 model accommodates a broader range of radii for a given mass, whereas the FSU model is, in general, more compact. 
A portion of the DD2+$K^-$ mass–radius posterior lies within the 68\% CIs of the mass-radius posterior (by both Riley and Miller) from PSR J0740+6620 NICER data. In contrast, for FSU+$K^-$, the entire posterior falls within the 68\% CI region from Riley's analysis for PSR J0740+6620 NICER data (dark pink).
\begin{figure*}
    \centering
    \includegraphics[width=\linewidth]{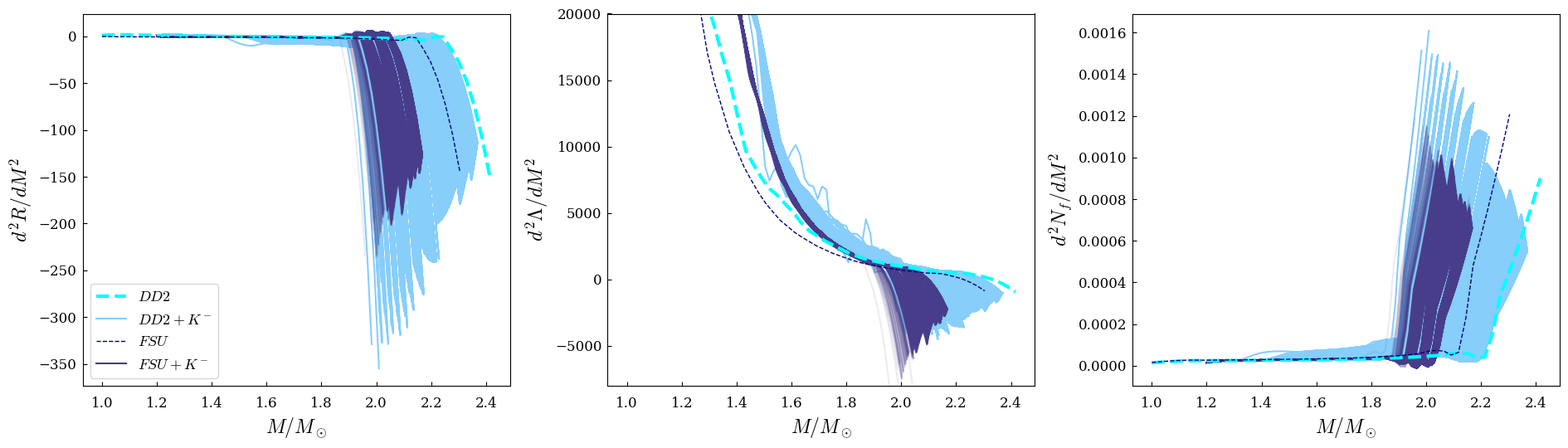}
    \caption{Variation of second-order derivatives: $\frac{d^2R}{dM^2}$, $\frac{d^2 \Lambda}{dM^2}$, and $\frac{d^2 N_f}{dM^2}$ for all the EOSs in this work.}
    \label{fig:deriv2_3panel}
\end{figure*}

The lower mass stars are composed only of nucleons and leptons, and the corresponding portions of mass–radius posteriors for both models overlap with the NICER constraints for PSR J0030$+$0451. However, only the FSU results are consistent with the mass-radius constraints from the GW170817 event.
Interestingly, the radius has the largest spread at $\sim2~M_\odot$ for the DD2+$K^-$ case, varying from $11.51$ to $13.41$ km. On the other hand, the FSU+$K^-$ case shows a minimal spread for the same mass, varying between a minimum of $12.19$ km and a maximum of $11.79$ km. This is a result of the observational constraints that we impose during the Bayesian inference, which forces the EOSs with antikaon condensates to support masses beyond $2~M_\odot$.

Finally the box in the mass-radius plane marks the boundary delineated by, $M_{\rm max}=2.25^{+0.08}_{-0.07}~M_\odot$ (68.3\% CI) and the corresponding $R_{\rm max}=11.90^{+0.63}_{-0.60}$ km as reported in Ref. \cite{Fan2024}.
We observe that 60\% of EOSs within the DD2 model and none of the EOSs within the FSU model predict maximum masses that satisfy those bounds. The distribution of the antikaon potential values associated with the fraction of EOSs within the DD2 model satisfying the bounds is plotted in Fig. \ref{fig:U_K_PDF}. Also, neither of the nucleonic models allows a maximum mass within the bounds. 

The horizontal bar in the middle panel of Fig. \ref{fig:MRLamNu} marks the GW170817 constraint on the tidal deformability of a $1.36$ $M_\odot$ NS.
For both models, neither the nucleon-only case nor the $K^-$-inclusive cases predict tidal deformabilities that satisfy those bounds.
The FSU+$K^-$-inclusive EOS barely touches the upper limit of the allowed range.
As a matter of fact, the core of a 1.36 $M_\odot$ NS in either of our models is not dense enough for $K^-$ condensation. Hence, it has no effect on putting the non-nucleonic EOSs within the tidal deformability bounds inferred from GW170817 data.

In the right panel of Fig. \ref{fig:MRLamNu}, the stellar mass is shown as a function of the $f-$ mode frequency $\nu_f$, the oscillations due to non-radial perturbations for the two RMF models.
The appearance of antikaons leads to a noticeable softening of the EOS, which in turn is reflected in the shift of the oscillation frequencies to higher values compared to those for nucleonic baseline or a lower $U_K$.
For DD2+$K^-$, the spread in $\nu_f$ peaks around $M \sim 2M_\odot$, while FSU+$K^-$ shows a comparatively narrower and more uniform spread across the mass range -- 
similar trend as in the radii of $2~M_\odot$ neutron stars.
This highlights the sensitivity of the $f-$mode spectrum to the onset of exotic degrees of freedom and their role in modifying the global stellar structure.
A fully general relativistic formalism for determining the oscillation frequencies allows us to evaluate the damping time $\tau_f$ corresponding to the $f-$mode oscillations.
Fig. \ref{fig:tau_f} demonstrates that $\tau_f$ systematically decreases once $K^-$ are introduced.
These results, taken together, establish a coherent picture: antikaon condensation not only increases the $f-$mode frequency but also accelerates the damping, with the effects most pronounced in the $2~M_\odot$ regime. 

Next, we look for interrelations between NS observables like $M$, $R$, $\Lambda$, $\nu_f$ and $\tau_f$ with the optical potential $U_K$ and the threshold density $\rho_{thres}$ through Pearson correlation coefficient. 
The correlation coefficients in Figs.\ref{fig:corr_plot_DD2} and \ref{fig:corr_plot_FSU} help us compare the interrelations in the presence of $K^-$ condensates for the two nucleonic EOSs with different underlying stiffness.
In both models, the antikaon optical potential $U_K$ is almost perfectly correlated with the maximum mass $M_{max}$ and the threshold density for $K^-$ condensation $n_{\mathrm{thres}}$ ($r \gtrsim 0.99$), and strongly anti-correlated ($r < -0.8$) with the tidal deformability of the maximum mass NS ($\Lambda_{\max}$).
Interestingly, the potential shows a completely different behaviour with respect to the radii of the maximum masses ($R_{max}$) in the two sets: moderately strong anti-correlation for the FSU+$K^-$ set and strong positive correlation for the DD2+$K^-$ set. The two completely different trends in variation of $M_{max}$ with the corresponding $R_{max}$ with $U_K$ in the mass-radius plane (see Fig. 3a) confirm this observation.
The real part of the fundamental quasinormal mode frequency, $\nu_f$, which is known to be inversely proportional to the radius, follows suit: $\nu_f$ for the maximum masses shows a strong positive correlation ($r \approx 0.8$) for the FSU+$K^-$ set and an equally strong anti-correlation for the DD2+$K^-$ set.
On average, $R_{max}$ increases with $M_{max}$ for the DD2+$K^-$ set, signalling a stiffening of the EOSs at high densities with increasing potential. On the other hand, the same decreases with $M_{max}$ for the FSU+$K^-$ set, implying a softening of the EOSs at high densities with increasing $U_K$.
In both cases, $R_{2.0}$ and $\Lambda_{2.0}$ are nearly perfectly correlated ($r \approx 0.99$) and anti-correlated with the fundamental mode frequency $\nu_{f,2.0}$, underscoring the tight coupling between stellar compactness, tidal response, and oscillation properties.
In essence, these correlations show that in a relatively soft EOS like FSU, $K^-$ condensation mainly shrinks massive stars and reduces their tidal deformability, while in a stiff EOS like DD2 it can instead slightly enlarge massive stars, highlighting that the interplay between exotic matter and stiffness can reverse the trend in radius-related properties.

\subsection{Curvature and derivatives with respect to stellar mass}
It has been shown that information from multiple points in the mass–radius plane, obtained from observations, can be utilised to constrain the NS matter EOS. For instance, Marcio Ferreira et al. \cite{Ferreira_PRD2024} attempted to constrain NS matter from the slope of the mass-radius curves at a given mass or more than one mass. Bauswein et. al. \cite{Bauswein_curvature_2025} on the other hand show that the slope of the $R(M)$ curve or that of $\lambda(M)$ and $\Lambda(M)= \lambda(M)/M^5$ 
at a fixed mass is not very conclusive on its own, as both the nucleonic and exotic models overlap with each other. However, the imprint of an exotic phase becomes distinct and clear if one considers the second-order derivative.
We also find that the slopes of $R(M)$, $\lambda(M)$, and $\Lambda(M)=\lambda/M^5$ in Fig. \ref{fig:first_deriv} overlap for nucleonic and exotic EOSs, making them inconclusive. 
In contrast, the second-order derivatives in Fig. \ref{fig:deriv2_3panel} show pronounced differences, clearly revealing the imprint of $K^-$ condensates as in Ref. \cite{Bauswein_curvature_2025} for hyperons. 
A similar behaviour is also observed for a scaled version of $f-$mode frequency, $N_f = \nu_f M^2$ (third panel of Fig. \ref{fig:deriv2_3panel}) in our analysis, underscoring the diagnostic power of higher-order derivatives for identifying EOS softening. 

Following Bauswein et al. \cite{Bauswein_curvature_2025}, we plot the variation of $\lambda$ along with that of the first-order and second-order derivative of $\lambda$ with respect to $M$ in Fig. \ref{fig:deriv_lambda}. The impact of antikaon condensation is much more pronounced for the second-order derivative compared to the first-order derivative.

\begin{figure}
    \centering
    \includegraphics[width=\linewidth]{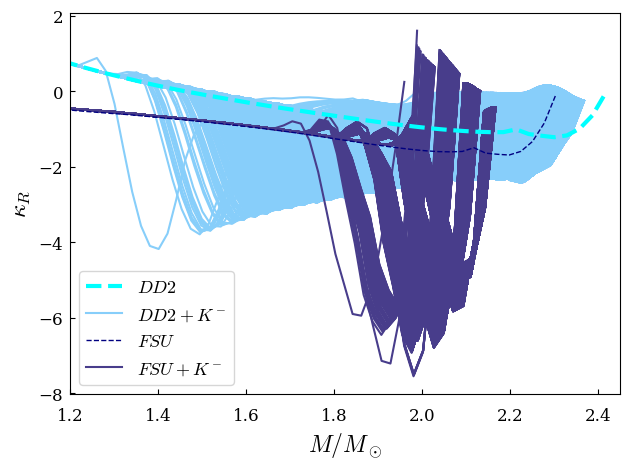}
    \caption{Plot of $\kappa_R$, the curvature of a curve $R(M)$ in the mas-radius plane for all the EOSs in this work}
    \label{fig:kappa_R}
\end{figure}

Bauswein et al. define curvature as \(\kappa_R =\frac{d^2 R}{dM^2} { (1 + (dR/dM)^2)^{-3/2} }\). In geometric units (G=c=1), the dimension of $\kappa_R$ is the inverse of length. 
The stellar mass at which $\kappa_R$ begins to decline significantly roughly marks the onset of $K^-$ production, as distinctly evident in Fig. \ref{fig:kappa_R}.
This implies that measuring $\kappa_R$ in the mass window $1.4~M_\odot \lesssim M \lesssim 2.2~M_\odot$ can help discriminate EOSs with exotic degrees of freedom from purely nucleonic ones. 
In particular, our nucleonic models do not reach curvatures smaller than about $\kappa_R \approx -1.5$. This trend is much more prominent for the FSU model, where $\kappa_R$ can reach as low as -7.
In this work, we find that the deviation of the mass–radius curvature from the nucleonic case, driven by antikaon condensation, underlies this behaviour. Moreover, the trends observed in the curvature of $R(M)$ and in the second derivatives of $R(M)$, $\Lambda(M)$, and $\lambda(M)$ for EOSs with antikaon condensation closely resemble those seen in EOSs with hyperons, as reported by Bauswein et al. Thus, these quantities alone do not provide a clear means to distinguish whether the EOS softening originates from antikaons or hyperons.

\section{Summary and Conclusions}
\label{sec:Conclusions}
In this work, we investigated the role of $K^-$ condensation in neutron star (NS) matter within the relativistic mean-field (RMF) framework using a Bayesian inference framework to put multimessenger (NICER and LIGO/Virgo) constraints on the equations of state (EOSs).
We explored a wide range of optical potential depths $U_K$ ($-60$ MeV to $-180$ MeV) to model antikaon condensation with two representative nucleonic EOSs --- a stiff (DD2 model) and soft (FSU model), and examined their impact on NS structure ($M$, $R$), tidal deformability ($\Lambda$), $f-$mode oscillation frequency ($\nu_f$) and the corresponding damping time ($\tau_f$).

Our analysis shows that deeper $K^-$ potentials trigger an earlier onset of kaon condensation, significantly softening the EOS and lowering the maximum supported mass. Bayesian posteriors favour $U_K = -104.72^{+13.82}_{-12.48}$ MeV with the associated 68\% credible interval (CI) for DD2 and a narrower band with $U_K = -66.46^{+2.47}_{-3.42}$ MeV for FSU when the observational bounds listed in Sec. \ref{sec:Bayes} are applied. The presence of antikaons alters the particle composition of NS cores and reduces the maximum masses to lie within $2.00$--$2.37$ $M_\odot$ (DD2) and $2.00$--$2.17$ $M_\odot$ (FSU), with corresponding radii within $11.08-12.35$ km.

While the slopes of the mass--radius and tidal deformability curves overlap between nucleonic and exotic EOSs, we found that \emph{their second derivatives and the curvature of the $R(M)$ curve} provide clearer diagnostics of EOS softening due to antikaon condensation. Similarly, the $f-$mode oscillation frequency and damping time exhibit measurable signatures: kaon-rich stars show higher oscillation frequencies and shorter damping times, especially near $2~M_\odot$. Correlation analyses reveal that the antikaon potential is tightly linked with maximum mass, threshold density, tidal deformability, and $f-$mode properties, reflecting the strong interplay between exotic matter and EOS stiffness.

However, the analysis of derivatives and curvature do not break the degeneracy between softening of EOS due to appearance of antikaons and hyperons at high densities. In fact, as pointed out in Bauswein et al. one may expect a similar behavior of the curvature and the relevant derivatives due to hadron-quark phase transition as well. Future works can attempt to resolve this issue by a systematic analysis of large sets of EOSs with various exotic degrees of freedom. We achieve better constraints on $U_K$ for the DD2 case by imposing strict bounds from Fan et al. \cite{Fan2024} for the maximum mass for a given $U_K$ and the corresponding radius. This shows that incorporating such a bound derived from merger simulation may be crucial for future EOS inferences resulting in tighter constraints on the properties of possible exotic phases of dense matter.

Overall, our results highlight that higher-order diagnostics---curvature of global NS properties and $f-$mode oscillations---offer promising probes of exotic physics at supranuclear densities and will become increasingly relevant as multimessenger observations reach greater precision. Future gravitational-wave detectors like the Einstein Telescope \cite{ET_WhitePaper2020} and Cosmic Explorer \cite{CE_WhitePaper2022}, combined with NICER measurements, could exploit these diagnostics to probe exotic matter inside the NS core.

\section*{Acknowledgements}
SB gratefully acknowledges the Department of Science \& Technology, India, for supporting the work via project no: CRG/2022/008360. DGR thanks Arya Raj and Anagh Venneti for their help in generating the equations of state used in this paper. The authors are indebted to Tuhin Malik for fruitful discussions and suggestions during the preparation of the manuscript.   

%% References
\bibliographystyle{elsarticle-num} 
\bibliography{refs_3rd_paper}

\begin{thebibliography}{10}
\expandafter\ifx\csname url\endcsname\relax
  \def\url#1{\texttt{#1}}\fi
\expandafter\ifx\csname urlprefix\endcsname\relax\def\urlprefix{URL }\fi
\expandafter\ifx\csname href\endcsname\relax
  \def\href#1#2{#2} \def\path#1{#1}\fi

\bibitem{glendenning}
N.~K. Glendenning, Springer, 1997.

\bibitem{Malik_ApJ_2021}
T.~Malik, S.~Banik, D.~Bandyopadhyay, Astrophys. J. 910 (2021) 96.

\bibitem{Banik_PRC2001}
S.~Banik, D.~Bandyopadhyay, Phys. Rev. C 63 (2001) 035802.

\bibitem{Friedman1999}
E.~Friedman, A.~Gal, Phys. Lett. B 459 (1999) 43.

\bibitem{Friedman2007}
E.~Friedman, A.~Gal, J.~Mareš, A.~Cieplý, Phys. Rev. C 76 (2007) 014008.

\bibitem{Bazzi2012}
M.~Bazzi, G.~Beer, L.~Bombelli, et~al., Nuc. Phys. A 881 (2012) 88–97.

\bibitem{KEKE228_Iwasaki1995}
M.~Iwasaki, K.~Bartlett, G.~Beer, et~al., Nuc. Phys. A 585 (1995) 239–246.

\bibitem{DEAR_Beer2005}
G.~Beer, A.~M. Bragadireanu, M.~Cargnelli, et~al., Phys. Rev. Lett. 94 (2005) 212302.

\bibitem{Sirghi2024}
F.~Sirghi, M.~Iliescu, L.~Abbene, et~al., J. Instru. 19 (2024) P11006.

\bibitem{Raaijmakers2021}
G.~Raaijmakers, S.~K. Greif, K.~Hebeler, et~al., Astrophys. J. Lett. 918 (2021) L29.

\bibitem{Rutherford2024}
N.~Rutherford, M.~Mendes, I.~Svensson, et~al., Astrophys. J. Lett. 971 (2024) L19.

\bibitem{Margalit2017}
B.~Margalit, B.~D. Metzger, Astrophys. J. Lett. 850 (2017) L19.

\bibitem{Radice2018}
D.~Radice, A.~Perego, F.~Zappa, et~al., Astrophys. J. Lett. 852 (2018) L29.

\bibitem{Tews_PRC2018}
I.~Tews, J.~Margueron, S.~Reddy, Phys. Rev. C 98 (2018) 045804.

\bibitem{Riley_2019}
T.~E. Riley, A.~L. Watts, S.~Bogdanov, et~al., Astrophys. J. Lett. 887 (2019) L21.

\bibitem{Miller_2019}
M.~C. Miller, F.~K. Lamb, A.~J. Dittmann, et~al., Astrophys. J. Lett. 887 (2019) L24.

\bibitem{Riley_2021}
T.~E. Riley, A.~L. Watts, P.~S. Ray, et~al., Astrophys. J. Lett. 918 (2021) L27.

\bibitem{Choudhury_2024}
D.~Choudhury, T.~Salmi, S.~Vinciguerra, et~al., Astrophys. J. Lett. 971 (2024) L20.

\bibitem{LIGOScientific:2018cki}
B.~P. Abbott, R.~Abbott, T.~D. Abbott, et~al., Phys. Rev. Lett. 121 (2018) 161101.

\bibitem{Annala_PRL2018}
E.~Annala, T.~Gorda, A.~Kurkela, et~al., Phys. Rev. Lett. 120 (2018) 172703.

\bibitem{Chatterjee_EPJA2016}
D.~Chatterjee, I.~Vidaña, Eur. Phys. J. A 52 (2016) 29.

\bibitem{Tolos_PPNP2020}
L.~Tolos, L.~Fabbietti, Prog. Part. and Nuc. Phys. 112 (2020) 103770.

\bibitem{Annala:2019puf}
E.~Annala, T.~Gorda, A.~Kurkela, et~al., Nat. Phys. 16 (2020) 907.

\bibitem{SunLattimer2025}
B.~Sun, J.~M. Lattimer, Astrophys. J. 984 (2025) 30.

\bibitem{Ferreira_PRD2024}
M.~Ferreira, C.~Provid\^encia, Physical Review D 110 (2024) 063018.

\bibitem{Bauswein_curvature_2025}
A.~Bauswein, A.~Nikolaidis, G.~Lioutas, et~al. (2025).
\newblock \href{https://arxiv.org/abs/2507.10372}{[link]}.
\newline\urlprefix\url{https://arxiv.org/abs/2507.10372}

\bibitem{ZhaoLattimer2022}
T.~Zhao, J.~M. Lattimer, Phys. Rev. D 106 (2022) 123002.

\bibitem{Lindblom_1983}
L.~Lindblom, S.~L. Detweiler, Astrophys. J. Supp. Series 53 (1983) 73.

\bibitem{Andersson_1998}
N.~Andersson, K.~D. Kokkotas, Mon. Not. Roy. Astron. Soc. 299 (1998) 1059.

\bibitem{Pradhan_PRC2022}
B.~K. Pradhan, D.~Chatterjee, M.~Lanoye, et~al., Phys. Rev. C 106 (2022) 015805.

\bibitem{Sotani_PRD2011}
H.~Sotani, N.~Yasutake, T.~Maruyama, et~al., Phys. Rev. D 83 (2011) 024014.

\bibitem{Thakur_PRD2024}
P.~Thakur, S.~Chatterjee, K.~K. Nath, et~al., Phys. Rev. D 110 (2024) 103045.

\bibitem{Kumar_EPJC2024}
A.~Kumar, M.~K. Ghosh, P.~Thakur, V.~B. Thapa, K.~K. Nath, M.~Sinha, Eur. Phys. J. C 84 (2024) 692.

\bibitem{Kumar_JCAP2023}
D.~Kumar, H.~Mishra, T.~Malik, JCAP 2023 (2023) 015.

\bibitem{CE_WhitePaper2022}
E.~D. Hall, Galaxies 10 (2022) 90.

\bibitem{ET_WhitePaper2020}
M.~Maggiore, C.~V.~D. Broeck, N.~Bartolo, et~al., JCAP 2020 (2020) 050.

\bibitem{Hempel_NucPhysA2010}
M.~Hempel, J.~Schaffner-Bielich, Nuc. Phys. A 837 (2010) 210.

\bibitem{Bunta_PRC2003}
J.~K. Bunta, S.~Gmuca, Phys. Rev. C 68 (2003) 054318.

\bibitem{Nandi_2019}
R.~Nandi, P.~Char, S.~Pal, Phys. Rev. C 99 (2019) 052802.

\bibitem{Banik_thirdFAMILY2001}
S.~Banik, D.~Bandyopadhyay, Phys. Rev. C 64 (2001) 055805.

\bibitem{BanikChar_PRC2014}
P.~Char, S.~Banik, Phys. Rev. C 90 (2014) 015801.

\bibitem{Friedman1994}
E.~Friedman, A.~Gal, C.~Batty, Nuc. Phys. A 579 (1994) 518.

\bibitem{Tolos_PRC2006}
L.~Tolós, A.~Ramos, E.~Oset, Phys. Rev. C 74 (2006) 015203.

\bibitem{Parmar2024}
V.~Parmar, V.~B. Thapa, A.~Kumar, et~al., Phys. Rev. C 110 (2024) 045804.

\bibitem{Kheto2023}
A.~Kheto, P.~Char, Eur. Phys. J. A 59 (2023) 201.

\bibitem{Postnikov2010}
S.~Postnikov, M.~Prakash, J.~M. Lattimer, Phys. Rev. D 82 (2010) 024016.

\bibitem{GuhaRoy_ApJ2024}
D.~Guha~Roy, T.~Malik, S.~Bhattacharya, et~al., Astrophys. J 968 (2024) 124.

\bibitem{GuhaRoy_PLB2024}
D.~Guha~Roy, A.~Venneti, T.~Malik, et~al., Phys. Lett. B 859 (2024) 139128.

\bibitem{Miller:2019cac}
M.~C. Miller, F.~K. Lamb, et~al., Astrophys. J. Lett. 887 (2019) L24.

\bibitem{Riley:2021pdl}
T.~E. Riley, et~al., Astrophys. J. Lett. 918 (2021) L27.

\bibitem{Miller:2021qha}
M.~C. Miller, et~al., Astrophys. J. Lett. 918 (2021) L28.

\bibitem{Antoniadis2013}
J.~Antoniadis, P.~C.~C. Freire, N.~Wex, et~al., Science 340 (2013) 448.

\bibitem{Arzoumanian2018}
Z.~Arzoumanian, A.~Brazier, S.~Burke-Spolaor, et~al., Astrophys. J. Supp. 235 (2018) 37.

\bibitem{Rezzolla_Jiang_2023}
J.-L. Jiang, C.~Ecker, L.~Rezzolla, Astrophys. J. 949 (2023) 11.

\bibitem{TerrellScott1992}
G.~R. Terrell, D.~W. Scott, Ann. of Stat. 20 (1992) 1236.

\bibitem{Buchner_2014}
J.~Buchner, A.~Georgakakis, K.~Nandra, et~al., Astron. I. Astrophys. 564 (2014) A125.

\bibitem{Fan2024}
Y.-Z. Fan, M.-Z. Han, J.-L. Jiang, et~al., Phys. Rev. D 109 (2024) 043052.

\end{thebibliography}

\clearpage
\onecolumn
\appendix
\section{Supplementary Material}
The supplementary material contains three sections: the first one shows the variation of the damping times corresponding to the $f-$mode oscillations for neutron stars (NSs) modelled with the equations of state (EOSs) with antikaon ($K^-$) condensation, and for those with the two chosen nucleonic EOSs for comparison. The second section displays the correlation among various stellar observables calculated in this work for the EOSs with $K^-$ condensation. The last section shows the plots for the slopes or the first-order derivatives of the radius ($R$), dimensionless tidal deformability ($\Lambda$), and a scaled version of the $f-$mode oscillation frequency $\nu_f$ defined as $N_f = \nu_f M^2$ with respect to the stellar mass $M$. 
The same section ends with the plots for the derivatives of the quantity $\lambda$ defined as $\Lambda M^5$, for all the EOSs in this work. 
\subsection{Damping times associated with f-mode oscillation}
\label{app:tau_f}
\begin{figure}[ht]
    \centering
    \includegraphics[width=0.7\linewidth]{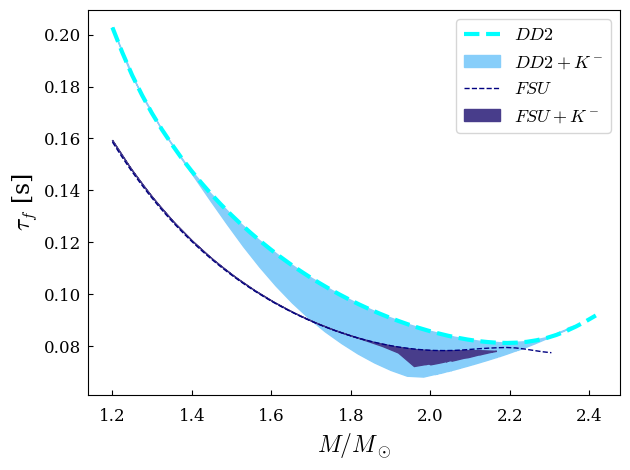}
    \caption{Plot of the damping time ($\tau_f$) corresponding to $f-$mode oscillations for the EOSs in this work.}
    \label{fig:tau_f}
\end{figure}
\clearpage
\subsection{Correlations among observables}
\label{app:corr_obs}
\begin{figure}[ht]
    \centering
    \includegraphics[width=0.7\linewidth]{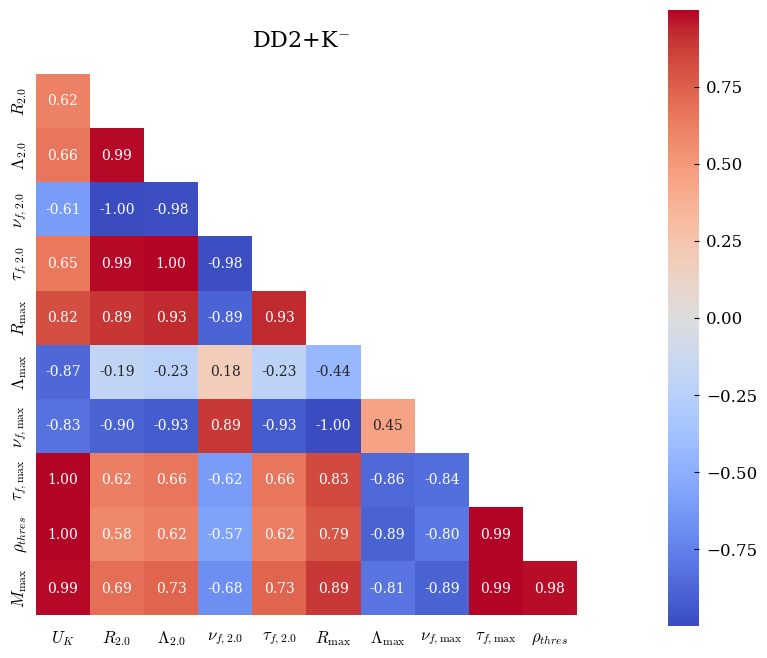}
    \caption{Pearson correlation coefficient between any pair among antikaon potential ($U_K$ [MeV]), radius of 1.4 $M_\odot$ and 2.0 $M_\odot$ neutron star ($R_{1.4}$ [km], $R_{2.0}$ [km]), dimensionless tidal deformability of a 1.4 $M_\odot$ and 2.0 $M_\odot$ ($\Lambda_{1.4}$, $\Lambda_{2.0}$), radius and dimensionless tidal deformability of the maximum mass neutron star for the DD2+$K^-$ equations of state corresponding to a $U_K$ value ($R_{max}$ [km], $\Lambda_{max}$).}
    \label{fig:corr_plot_DD2}
\end{figure}
\begin{figure}[ht]
    \centering
    \includegraphics[width=0.7\linewidth]{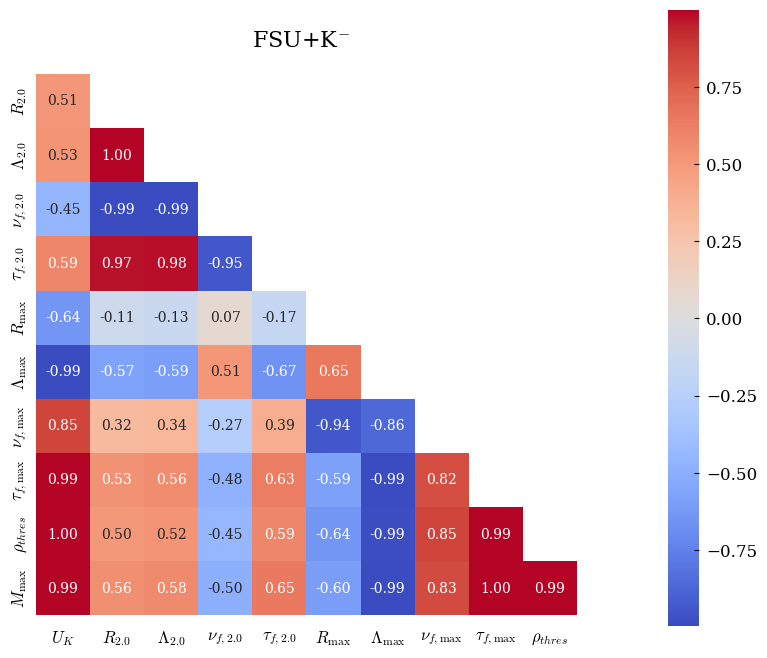}
    \caption{Same as Fig. A.7 but for FSU+$K^-$ equations of state.}
    \label{fig:corr_plot_FSU}
\end{figure}
\clearpage
\subsection{Derivatives of observables with respect to stellar mass}
\label{app:deriv}
\begin{figure*}[htp]
    \centering
    \begin{tabular}{ccc}
        \includegraphics[width=0.32\textwidth]{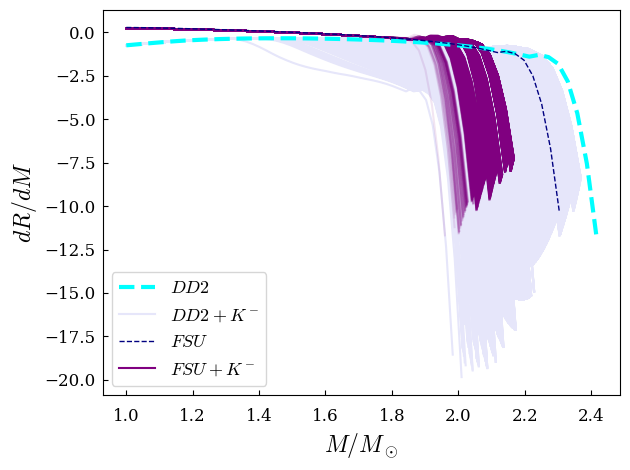} &
        \includegraphics[width=0.32\textwidth]{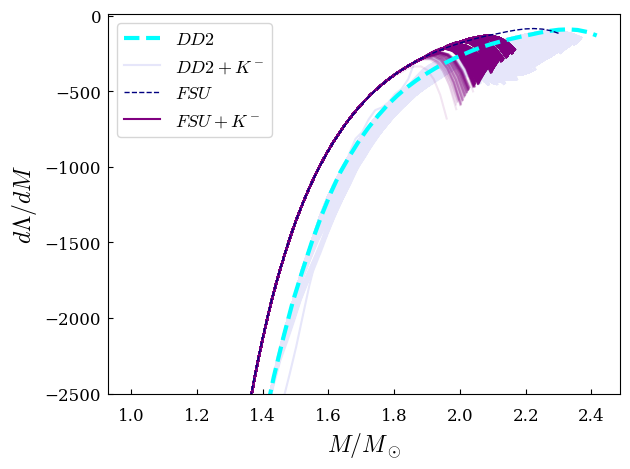} &
        \includegraphics[width=0.32\textwidth]{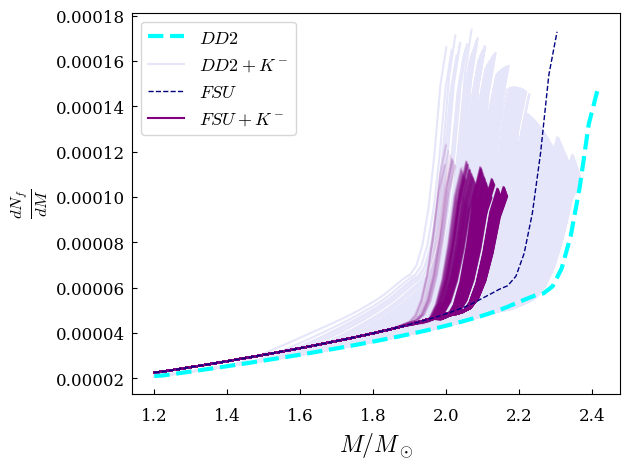} \\
    \end{tabular}
    \caption{\label{fig:first_deriv}Variation of first-order derivatives: $\frac{dR}{dM}$, $\frac{d \Lambda}{dM}$, and $\frac{d N_f}{dM}$ for all the EOSs in this work.}
\end{figure*}
\begin{figure*}[htp]
    \centering
    \begin{tabular}{ccc}
        \includegraphics[width=0.32\textwidth]{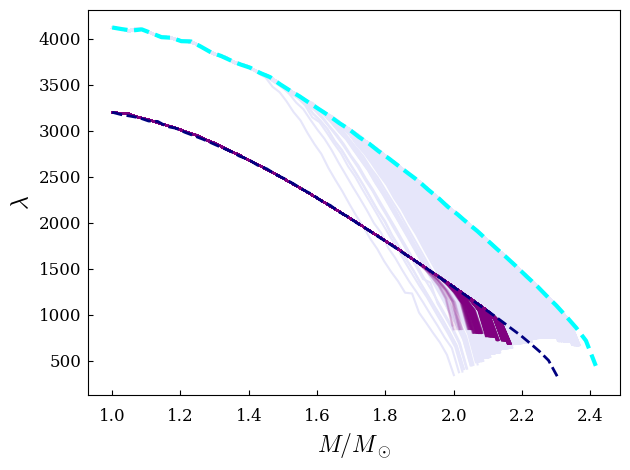} &
        \includegraphics[width=0.32\textwidth]{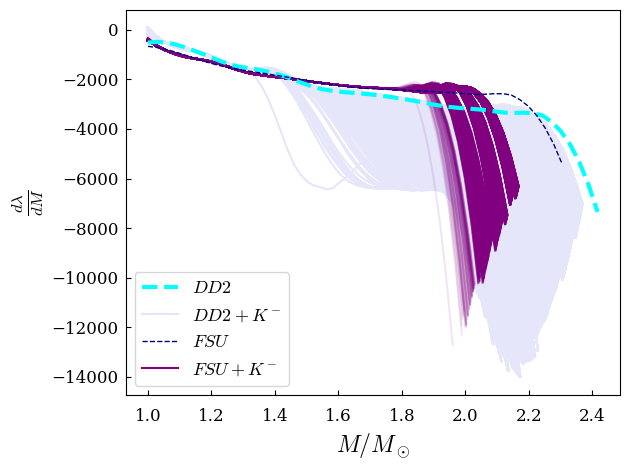} &
        \includegraphics[width=0.32\textwidth]{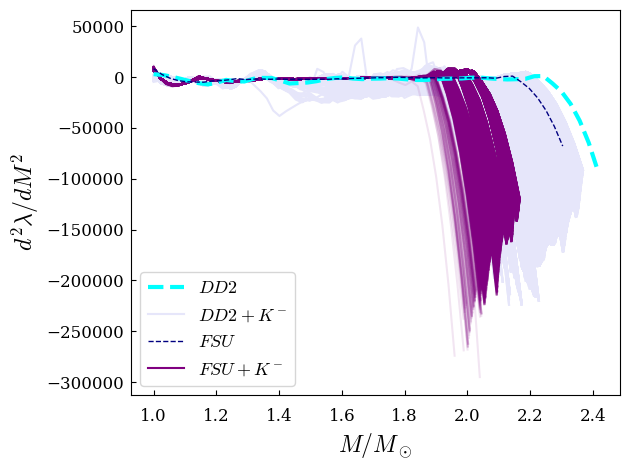} \\
    \end{tabular}
    \caption{\label{fig:deriv_lambda}Variation of $\lambda$, first-order derivative of $\lambda$ and second-order derivative of $\lambda$ with respect to mass $M$ for all the EOSs used in this work.}
\end{figure*}

\end{document}